# Simulation of Switching Converters Using Linear Capacitor Voltage and Inductor Current Prediction and Correction

Aleksandra Lekić, Vujo Drndarević
School of Electrical Engineering
Belgrade, Serbia
lekic.aleksandra@etf.rs, vujo@etf.rs

*Abstract* — **In this paper an algorithm for transient simulation of switching converters using prediction and correction to calculate duty ratio is proposed. It provides large signal simulation on the level of averaged currents and voltages in the circuit. Calculation of duty ratio using inductor current and capacitor voltage prediction and correction do not require their priori knowledge. Number of circuit solving per switching period is fixed and equal to two. Using this algorithm various of constant frequency regulated switching converters can be simulated. Due to predetermined circuit values convergence problems are avoided. This algorithm results in very fast and accurate large signal simulation.**

## I. INTRODUCTION

Transient simulation of power electronic circuits is often necessary in order to check its performance. Therefore, it has been frequently research topic [1]-[11]. However, it is a very demanding task [1] and variety of methods are developed employing transient analysis.

General purpose simulators are designed to be able to simulate wide range of electronic circuits. Therefore, while simulating power electronic circuits, which incorporate fast switching elements, sometimes have convergence problems. These problems are caused by Newton-Raphson iterative method for solving nonlinear equations that all general purpose simulators apply. In some cases convergence problems can be avoided by choosing smaller simulation time step. But there are cases when decreasing time step do not lead to solution and simulation cannot be performed.

Other types of simulators are simulators designed for power electronic systems. They can be sorted into two groups: piecewise-linear simulators and simulators based on averaging. Piecewise-linear simulators such as [2]-[4] have linear models for each operating segment of nonlinear element. Therefore, different simulators have different algorithms for evaluation in which segment is element operating at the specific time. Convergence problems are avoided, but determination of operating segments of the piecewise-linear elements is a newly introduced problem which can lead to problems similar to the convergence problems.

Simulators based on averaging are developed in order to avoid previously described simulation problems. The idea to perform averaged numerical simulation is in the fact that line frequency is much lower than the converter switching frequency $f_S \gg f_0$. If the averaged simulation is properly performed, provided results can give enough accurate signal spectra, RMS values, etc. The most popular algorithms based on averaging are state-space averaging [5] and switching cell averaging [6]. State-space averaging is based on writing state-space model for every state during one switching interval and their averaging [7]. Lot of research efforts are invested in improving this averaging method, like in [5] and [7], by introducing special multiplication matrices in order to get better spectral response for converter operating in discontinuous conduction mode. Other approach performs averaging on the level of switching cell [6]. Paper [8] introduces fourteen families of switching cells.

Some of known methods perform averaging based on exact solution of differential equations contained in the state-space models. This algorithm solves state-space equations linearized by the Newton-Raphson method which can lead to previously described convergence problems. It should be noted that performing transient of power electronic circuit by exact solving of differential equations is rather slow, which will be shown in the section V.

In this paper a new method for simulation of power electronic circuits based on circuit averaging is described. Averaging is done on the level of switching cell, named Cell A in [8], with predefined switching cell equations. These equations are easy to implement and due to them convergence problems and operating segment problems are avoided. As a result of switching cell averaging, assuming quasi steady state approximation and linear ripple approximation, inductor current and capacitor voltage ripple can be estimated. That provides possibility of construction of instantaneous waveforms. Equations that describe inductor currents and capacitor voltages are nonlinear and rely on the duty cycle value in the current switching interval. In order to avoid necessity for its a priori knowledge, predictor-corrector integration method is used. Predictor-corrector integration method is introduced in [9] for PWM controlled converters. This algorithm can be applied for current mode controlled converters such as peak limiting control as well. Predictor-corrector integration method applied in this paper reduces size of the system matrix to be solved by dividing it into two smaller matrices, one for converter and one for regulator, and also limits number of solutions to be performed by fixing it into two steps per period, first for prediction and the second for correction. As a result computational efforts are reduced.

Motivation for performing this research is to provide computationally fast simulator that can avoid convergence problems and can give information about transient in the case of regulation with "hard nonlinearities". By "hard nonlinearities" are considered all types of constant frequency controls, as a current peak limiting or duty ratio limiting. Nonlinearities got as a result of dynamics (losses, etc.) are not considered because they are negligible comparing to the hard ones.

Section II of this paper contains predefined switching cell equations. Predictor-corrector integration method is described in section III. In the section IV proposed algorithm is described and in the section V simulation results are given.

## II. SWITCHING CELL

In this section switching cell used in simulation is described. Switching cell is considered as a three terminal, two-port element described by the set of four equations. In this paper only basic switching cell is described. Thus, switching cell can be realized as a common one, operating both in continuous and discontinuous conduction modes, with unidirectional switch S, inductor and a diode D and bidirectional, that operates only in continuous conduction mode (CCM), with two bidirectional switches (state(S1) = ¬ state(S2)) and inductor. Using switching cells' types depicted in Fig. 1 all of the basic converters, buck, boost and buck-boost, can be simulated. These switching cells are described in [8] as type Cell A.

Averaged model of the switching cell operating both in continuous and discontinuous conduction mode is depicted in Fig. 2. It is assumed that during one switching interval, $nT_S \leq t < (n+1)T_S$, terminal voltages are constant and that $v_1 > 0\,\text{V}$ and $v_2 < 0\,\text{V}$. Considering previous assumption and using quasi steady state approximation, equations that describe inductor voltage and inductor current can be written. In the continuous conduction mode, switching interval is divided in two subintervals: when the switch S1 (switch S in Fig. 1b) is on, $nT_S \leq t < (n+d)T_S$ and when the switch S2 (diode D in Fig. 1b) is on, $(n+d)T_S \leq t < (n+1)T_S$. In the discontinuous conduction mode (DCM) there are three subintervals: when the switch S is on, $nT_S \leq t < (n+d)T_S$, when the diode D is on, $(n+d)T_S \leq t < (n+d+d_2)T_S$ and when the switch and a diode are off, $(n+d+d_2)T_S \leq t < (n+1)T_S$. Thus, equations for inductor voltage during one switching interval are as follows.

$$v_L = \begin{cases} v_1 & nT_S \leq t < (n+d)T_S \\ v_2 & (n+d)T_S \leq t < (n+d+d_2)T_S \\ 0 & (n+d+d_2)T_S \leq t < (n+1)T_S \end{cases} \quad (1)$$

Assuming that $i_{L0}$ is the inductor current at the beginning of the period, $i_{L0} = i_L(nT_S)$, $i_{L1} = i_L((n+d)T_S)$ is current at the moment $t = (n+d)T_S$ and $i_{L2} = i_L((n+d+d_2)T_S)$ at the end of the period, following equations can be derived:

$$i_{L1} = i_L((n+d)T_S) = i_{L0} + \frac{v_1}{L}dT_S, \quad (2)$$

$$i_{L2} = i_L((n+d+d_2)T_S) = i_{L1} + \frac{v_2}{L}d_2T_S. \quad (3)$$

In case of operating in CCM $i_{L2}$ has some positive value, but in the DCM $i_{L2} = 0$. In the second case during the third subinterval inductor current is constant (because $v_L = 0$) and $i_L = 0$. Thus $i_{L2}$ is the current at the end of the switching interval in DCM as well.

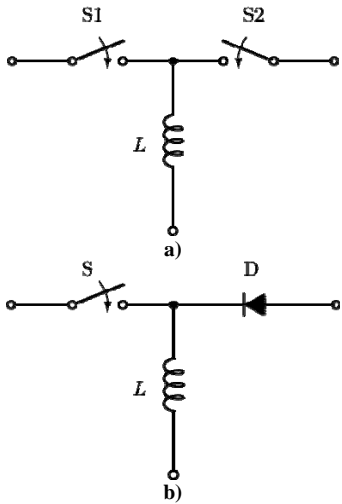

Figure 1. Types of basic switching cells: a) with bidirectional switches and inductor; b) with switch, diode and inductor.

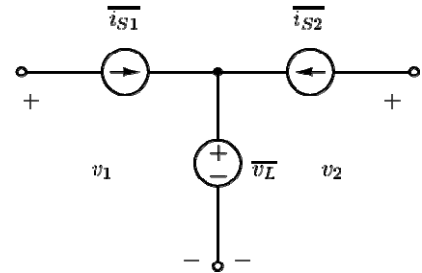

Figure 2. Averaged model of a switching cell.

According to the model from Fig. 2, equations used to describe switching cell during one switching period are:

$$\overline{i_{S1}} = d i_{L0} + \frac{d^2 G_L}{2} v_1 \quad (4)$$

$$\overline{i_{S2}} = d_2 i_{L0} + d d_2 G_L v_1 + \frac{d_2^2 G_L}{2} v_2 \quad (5)$$

$$v_{L1} = d v_1 \quad (6)$$

$$v_{L2} = d_2 v_2 \quad (7)$$

where $G_L = T_S / L$ and $d_2$ presents $d_2 = 1 - d$ in CCM and in DCM

$$d_2 = -\frac{v_1}{v_2} d . \quad (8)$$

Converter operates in DCM if $d + d_2 < 1$, otherwise operates in CCM. Equations (4)-(7) present switching cell model used for both CCM and DCM.

### III. INTEGRATION METHOD

In this section predictor-corrector method used for capacitor voltage and inductor current calculation is proposed. This method is necessary for evaluation of duty ratio value further used in solving averaged convertor model given with Eqs. (4)-(7) and equations for nodes in converter circuit obtained through modified nodal analysis [6]. Afterwards regulator construction is described.

#### A. Linear Capacitor Voltage and Inductor Current Prediction

State variables in switching converters are given in form of the capacitor voltage and switching cell (inductor) current. In order to determine duty ratio during current switching period, those values are previously predicted. Prediction is done according to Forward Euler numerical integration. Considering $nT_S \leq t < (n+1)T_S$ switching period, predicted averaged capacitor voltage value in discretized form is

$$v_C^P[n] = v_C[n-1] + \frac{T_S}{C} i_C[n-1] . \quad (9)$$

For switching cell current $i_{L1}$ at the moment $t = (n+d)T_S$ is predicted.

$$i_{L1}^P[n] = i_{L0}[n] + G_L d[n-1] v_1[n-1] \quad (10)$$

Current $i_{L0}$ presents inductor current at the end of previous switching period and thus, at the beginning of the current period already known from the previous time step. After performing prediction step, predicted duty ratio is calculated, $d^P[n]$. It is used to perform evaluation by solving converter circuit equations written as slightly optimized form of modified nodal analysis. This optimization is done by adding switching cell equations (4)-(7) and capacitor voltage equations provided by trapezoidal integration formula. Prediction and evaluation step is followed by correction of capacitor voltage and inductor current value according to trapezoidal integration formula.

$$v_C^C[n] = v_C[n-1] + \frac{T_S}{2C} \left( i_C[n-1] + i_C^P[n] \right) \quad (11)$$

$$i_{L1}^C[n] = i_{L0}[n] + \frac{G_L}{2} \left( d[n-1] v_1[n-1] + d^P[n] v_1^P[n] \right) \quad (12)$$

Superscript P is used for predicted values and values calculated during evaluation, and C for corrected. The method described is PECE predictor-corrector method.

#### B. Regulator Construction

Regulator can be designed as constant frequency PWM or current mode controlled (CMC). There are two ways for regulator realization. In the first case when there is no limiting applied, regulator is given as bilinear transformed transfer function. In that case there are coefficients that multiply current output value and for that matter predicted and in second step evaluated values are used. Another case of regulator is some kind of limiting. In case when duty ratio is limited, it has to be ensured that duty ratio value does not exceed given value. More complicated case is peak limiting control. According to [10] duty ratio can be obtained through following equation.

$$d = \frac{I_{ref} - i_{L0}}{I_{slope} + G_L v_1} \quad (13)$$

Equation (13) provides calculation of the duty ratio according to known peak current value $i_{L1} = I_{ref}$ at the moment $t = (n+d)T_S$. In the equation (13) $I_{ref}$ presents maximum allowed current value, while $I_{slope}$ presents amplitude of the PWM signal.

Regulator can be designed to incorporate few ways of controlling and more limitations [11]. In such cases duty ratio is calculated for each controlling part and the smallest value is taken for sequent converter circuit solving.

### IV. SIMULATION METHOD

During one switching period two sequential circuit solving are applied. First one employs prediction of the inductor current (switching cell current) and capacitor voltage and those values are used to calculate duty ratio, which is then used to perform evaluation of all circuit variables. After prediction step, inductor current and capacitor voltage values are corrected according to the integration formula described in the previous section, duty ratio is obtained and new evaluation is performed. To test the proposed algorithm simple program in the interpretative language Python 2.7.6 is written.

Regulated converter circuit is divided into two parts: converter and regulator. Converter consists of independent input source, one switching cell and a number of passive and active elements. Regulator is described in a discretized form applying bilinear transformation to desired transfer function. Discretization is done using sampling time equal to switching period $T_S$. Used sampling time is valid under already taken assumption $f_S \gg f_0$. It is possible to add some hard nonlinearities by limiting desired values inside regulator. In this paper only constant switching frequency converters are considered: peak limiting current mode control and PWM regulation. Feedback between converter and regulator is accomplished through predicted/corrected inductor current and capacitor voltage values, and between regulator and converter through the information about current duty ratio.

After finished averaged simulation ripple can be superimposed applying procedure described in [6]. Inductor current instantaneous waveform is constructed by connecting $i_{L0}$, $i_{L1}$ and $i_{L2}$ with strait line segments. Waveforms of capacitor voltages are constructed as in [6].

## V. SIMULATION EXAMPLES

To illustrate application of the simulation method, two examples will be described and their simulation results will be presented. In order to compare computational speed of the proposed algorithm with other available simulators and algorithms, provided examples are tested using PETS [2] and PLECS Blockset version [3] and algorithm named EXACT, implemented as program written in Python 2.7.6.

EXACT algorithm is based on exact solving of differential equations, using eigenvalues and eigenvectors for calculation spectral components and writing equations for the state variables (inductor current and capacitor voltage). State equations and output equations are completely written using state-space matrices. Operating state change time is determined precisely using second order Newton-Raphson method described in [4]. Values of state-space variables are calculated in 1000 points per period.

In the Fig. 3 is depicted regulated buck converter used frequently as test example [2], [4], [11]. Buck regulator from Fig. 3 has soft-switching start, driver which applies duty ratio limitation to 0.85, inductor current limitation to 4 A and switching frequency $f_S = 20\,\text{kHz}$. Input voltage has step change from 20 V to 40 V at the time 100 ms. Circuit transient is simulated during first 20 ms and that simulation took 95.2841 ms real and 20 ms CPU time. Results of the simulation are shown in Fig. 4 where maximum, minimum and average inductor currents are presented, as well as constructed instantaneous inductor current and output voltage waveforms. In order to check transient due to the input voltage step change, circuit is simulated during 200 ms. The waveforms after step change are shown in the Fig. 5. That simulation lasted only 950.182 ms on PC equipped with Intel i7-3537U CPU run at 2.5 GHz.

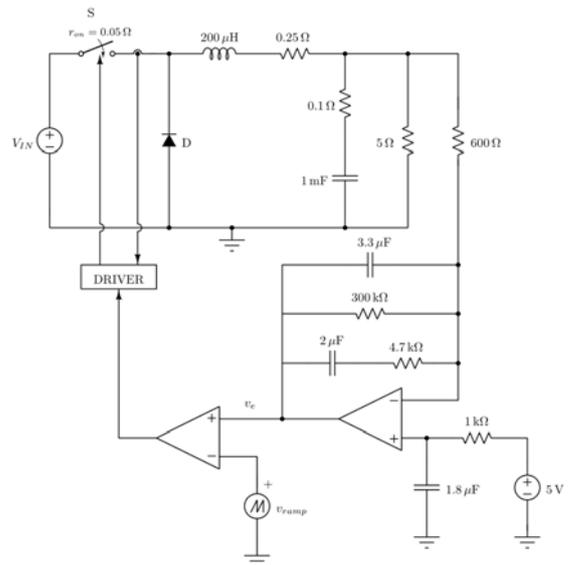

Figure 3. Regulated buck converter.

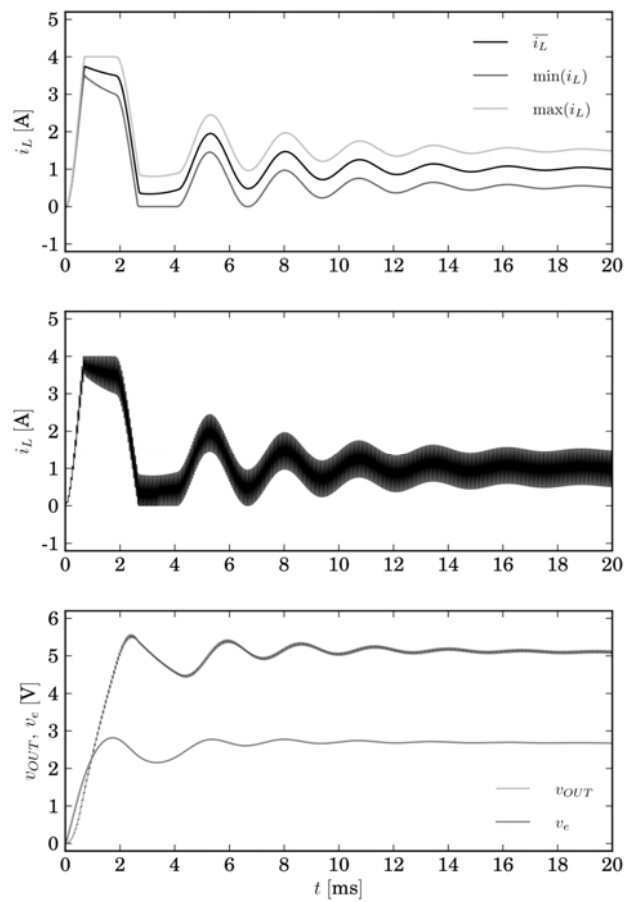

Figure 4. Simulated waveforms at start-up of regulated buck converter depicted in Fig. 3.

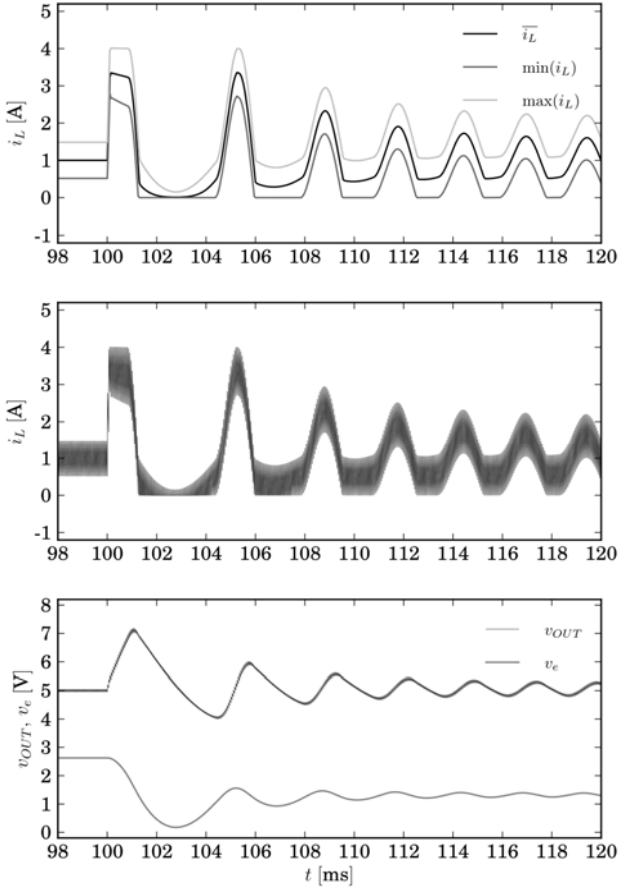

Figure 5. Simulated waveforms due to step change of input voltage from 20 V to 40 V of regulated buck converter depicted in Fig. 3.

In the table I simulation computational times for the presented algorithm, simulators PETS and PLECS and EXACT algorithm for described examples are presented.

Second example shown in Fig. 6 is also used as testing circuit in [11]. This example in lot of simulators produces multiple zero crossings. In cases when those zero crossings cannot be properly handled, simulator does not produce transient diagrams. That happens in case of PETS, where simulation cannot be performed. Similar case is with PLECS Blockset integrated with MATLAB Simulink. However, Standalone version of PLECS can simulate this example with enormously decreased precision and rather smaller computational speed, but it does not provide tool for calculation of simulation time. Proposed method does not have zero crossings due to precalculated values of voltages, currents and duty ratio, which are to be used to determine operating mode. Simulation is done for cases when filter is excluded from regulator and with filter included. Filter applied has two poles at 50 Hz line frequency, transfer function $H(s)$. It should be mentioned that the second case, when filter is applied, produces more zero crossings which makes it even more demanding circuit for simulation. In the Fig. 7 is shown start-up transient when filter is excluded and in Fig. 8 when the filter is included. Table I contains simulation time for the circuit of Fig. 6, as well.

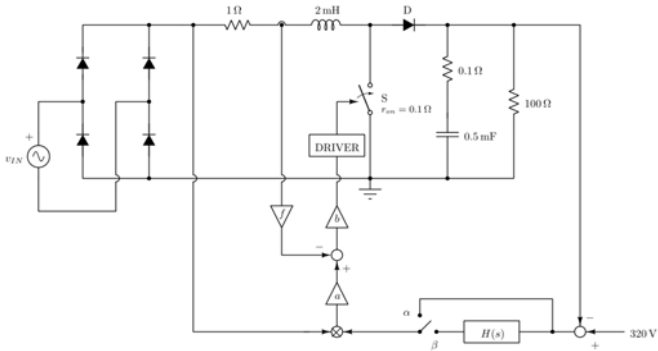

Figure 6. Switch mode rectifier. Parameters are:
$v_{IN} = 300 \text{ V} \sin(2\pi f_0)$, $f_0 = 50 \text{ Hz}$, $f_S = 20 \text{ kHz}$,
$a = 2 \times 10^{-4} \text{ V}^{-1}$, $b = 30$, $f = 0.2 \text{ V/A}$, $H(s) = \dfrac{1}{(1 + s/(2\pi f_0))^2}$.

TABLE I. SIMULATION TIME COMPARISON

| Simulation method | Proposed method | EXACT | PETS | PLECS |
|---|---|---|---|---|
| Simulation time for 20ms transient, circuit in Fig. 3 | 95.28 ms | 28.68 s | 1.87 s | 2.43 s |
| Simulation time for 200 ms transient, circuit in Fig. 3 | 950.18 ms | 394.93 s | 10.02 s | 22.21 s |
| Switch mode rectifier with filter excluded, Fig. 6, 800 ms transient | 396.59 ms | 163.35 s | - | 180.06 s |
| Switch mode rectifier with filter included, Fig. 6, 800 ms transient | 413.68 ms | 42.05 s | - | - |

Proposed algorithm provides computationally fast transient simulation. From the table I can be seen that all given examples require smaller simulation time using proposed algorithm than with PETS, PLECS and EXACT on PC equipped with Intel i7-3537U CPU run at 2.5 GHz. Proposed algorithm, PLECS and EXACT simulations run on 64bit Windows 7. PETS is running only on 32bit operating systems and therefore Windows XP is used. Accuracy of the algorithm described in this paper has been estimated comparing to EXACT. After superimposing ripple to average values according to [6] absolute error is up to 5 %. For the purpose of fast simulation used to determine circuit transient, this is very good accuracy.

## VI. CONCLUSION

An algorithm for simulation switching converters on the level of averaged values of voltages and currents is presented in this paper. This algorithm relies on averaging on the level of switching cell and applies regulation using prediction and correction. Algorithm is implemented by the use of interpretative programming language Python 2.6.7. Computational time is very small even though it is implemented in interpretative programming language. Reason for that is fixed number of computations per period, one for prediction step and one for correction, and maximally increased simulation time step equal to switching period. That is done under assumption that switching frequency is much higher than the line frequency. Simulation provides averaged values of voltages and currents in the circuit that can be further processed to construct instantaneous waveforms by superimposing ripple.

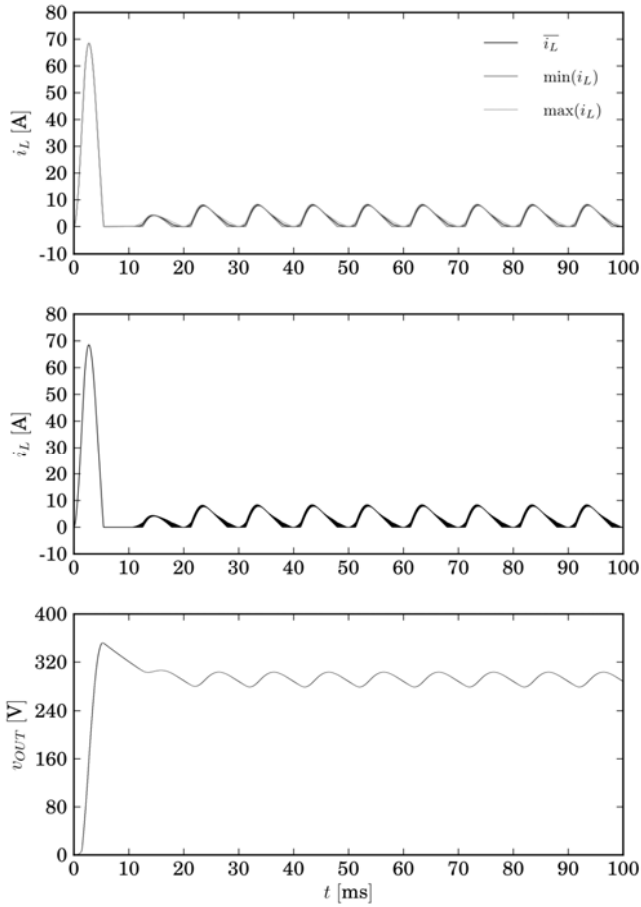

Figure 7. Simulated waveforms at start-up of rectifier from Fig. 6, filter excluded.

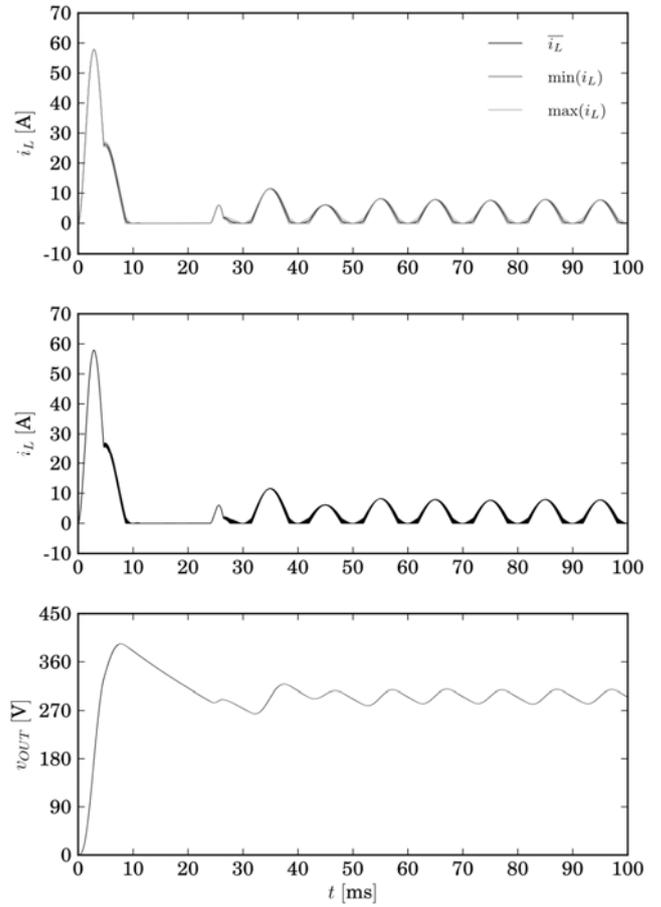

Figure 8. Simulated waveforms at start-up of rectifier from Fig. 6, filter included.

Proposed algorithm is tested on PC equipped with the Intel i7-3537U CPU run at 2.5 GHz under Windows 7 operating system. It shows good level of accuracy and presents great advantages of computational speed.